# Where Journalism Silenced Voices: Exploring Discrimination in the Representation of Indigenous Communities in Bangladesh

ABHIJIT PAUL, IIT, University of Dhaka, Bangladesh
ADITY KHISA, IIT, University of Dhaka, Bangladesh
ZARIF MASUD, Toronto Metropolitan University, Canada
SHARIF MD. ABDULLAH, IIT, University of Dhaka, Bangladesh
AHMEDUL KABIR, IIT, University of Dhaka, Bangladesh
SHEBUTI RAYANA, Math/CIS, SUNY at Old Westbury, USA

In this paper, we examine the intersections of indigeneity, media representation, social equity, and technology in shaping perceptions of indigenous communities in Bangladesh. Despite a population exceeding 1.6 million, indigenous people in Bangladesh often feel invisible in mainstream media. Using a mixed-methods approach, we combine quantitative analysis of media data with qualitative insights from focus group discussions (FGD). First, we identify a total of 4,893 indigenous-related articles from our initial dataset of 2.2 million newspaper articles, using a combination of keyword-based filtering and prompting a large language model (LLM), achieving 77% accuracy and an F1-score of 81.9%. From manually inspecting 3 prominent Bangla newspapers, we identify 15 genres that we use as our topics for semi-supervised topic modeling using CorEx. Results show indigenous news articles have higher representation of culture and entertainment (19%, 10% higher than general news articles), and a disproportionate focus on conflict and protest (9%, 7% higher than general news). On the other hand, sentiment analysis reveals that 57% of articles on indigenous topics carry a negative tone, compared to 27% for non-indigenous related news. Drawing from communication studies, we further analyze framing (using LLM, agreement score 58%), priming (Wordcloud analysis), and agenda-setting (frequency of themes) to support the case for discrimination in representation of indigenous news coverage. For the qualitative part of our analysis, we facilitated FGD, where participants further validated these findings. Participants unanimously expressed their feeling of being under-represented, and that critical issues affecting their communities (such as education, healthcare, and land rights) are systematically marginalized in news media coverage. By highlighting 8 cases of discrimination and media misrepresentation that were frequently mentioned by participants in the FGD, this study emphasizes the urgent need for more equitable media practices that accurately reflect the experiences and struggles of marginalized communities. Our research contributes to the discourse on indigeneity and technology by using both quantitative and qualitative methods to find evidence of discrimination in representation, advocating for more inclusive media narratives.

CCS Concepts: • **Human-centered computing** → **Cultural factors in computing**; *User studies*; • **Social and professional topics** → Discrimination; • **Applied computing** → Media studies.

Additional Key Words and Phrases: Discrimination of Representation, indigenous communities, media representation, Chittagong Hill Tracts, qualitative research

## 1 Introduction

Bangladesh is an ethnically diverse country with over 54 indigenous communities, speaking at least 35 languages. According to the 2022 census, the indigenous population numbers approximately 1,650,159 or 1% of the total population [18]. However, indigenous communities claim that their population is much higher than this [12]. Bangladesh has yet not adopted the UN Declaration on the Rights of Indigenous Peoples. Historically marginalized, these communities face multifaceted challenges, including severe human rights violations, land dispossession [68], and systemic misrepresentation in media narratives. Such issues are not merely local concerns; they reflect broader patterns of discrimination and inequality that resonate within global discussions on indigenous rights.

Most of these communities live in the Chittagong Hill Tracts (CHT), while others live in the *flatlands* (such as Mymensingh, Rajshahi, and Sylhet). CHT present a complex landscape marked by a rich cultural heritage intertwined with ongoing struggles for rights and recognition. Additionally, indigenous people living in flatlands alongside Bengali people also have their unique experiences and struggles. The United Nations Permanent Forum on Indigenous Issues (UNPFII) reported an excessive amount of militarization in CHT with arbitrary arrests, torture, and extrajudicial killings. UNPFII also reports human rights violation and rampant sexual harassment [68]. They are also marginalized in mainstream media, something we explore in great details in this paper.

It is worth noting that the state does not recognize these communities as "Indigenous". Prior to the International Day of the World's Indigenous Peoples in 2022, the Ministry of Information and Broadcasting of Bangladesh issued a circular directing all the electronic media of Bangladesh not to use the term "Indigenous" [1]. However, according to Computer Supported Cooperative Work (CSCW) 2023 workshop in indigeneity, the term "Indigenous" is not just a label of ethnic or geographical classification. It encapsulates communities that have historically been marginalized, bearing unique cultural, social, and historical narratives often overshadowed by dominant global narratives [9]. Moreover, United Nations Economic and Social Council (ECOSOC) [69], and various prominent work in literature [46, 60] also use the term *Indigenous* to refer to these communities.

Despite the struggles of indigenous communities in Bangladesh, the broader Bengali population remains largely oblivious [58]. We hypothesize that a key reason for this is the under-representation of these issues in the media. Moreover, media coverage shapes dominant perspectives through stereotyping and under-representation which negatively impacts identity formation. Inspired by the approach in [50], where the authors discuss this phenomenon for the Asian diaspora, we use quantitative methods to first estimate how news media over-represents dominant perspectives (i.e. culture, conflict and protests), and then apply qualitative methods to explore how this over-representation negatively impacts indigenous people by perpetuating stereotypes and under-representing critical experiences.

Kojah et al. mentions "Empowerment Through Participation", suggest that indigenous communities are often studied without having control over the funding, framing, or the language used to describe their experiences [49]. They should be seen not just as subjects, but as active participants who can re-imagine how technology fits into their lives and shape its design. It is, therefore, crucial to center indigenous voices when conducting such studies. In the qualitative part of our analysis, we facilitate a focus group discussion (FGD) intended to empower and amplify indigenous voices. Moreover, one of the authors of the paper is a member of an indigenous community.

This research highlights the stark disparity between the pressing realities faced by indigenous communities and the narratives that dominate public discourse. Using various natural language processing (NLP) techniques like keyword-based filtering, topic modeling, and sentiment analysis, we show that topics such as conflict, protest, and culture are over-represented in news regarding indigenous communities, compared to general (non-indigenous related) news, and with a higher degree of negativity in tone. FGD corroborated these findings, with participants reporting that essential issues affecting their lives—such as education, healthcare, and land rights—are consistently overlooked in media coverage. By examining multiple instances of discrimination, we aim to shed light on the systematic erasure of indigenous experiences and the implications this has for societal understanding and policy action.

The contributions of this paper are as follows.

- Finding strong support for discrimination of representation of indigenous communities in Bangladesh through quantitative and qualitative analysis.
- Highlighting critical issues affecting indigenous communities that are often overlooked in media narratives.
- Examining the impacts of discrimination of representation on indigenous people.
- Offer actionable recommendations for media practitioners to enhance accurate representation.

With the newfound sense of freedom following the mass uprising in Bangladesh in August 2024 [2], we aim to draw attention to the discrimination of representation of indigenous communities in Bangladeshi media and how severely it impacts them by addressing issues of media representation, social equity, and the role of technology in shaping collective understanding. We explore how technology and media can perpetuate or challenge biases, impacting marginalized communities.

## 2 Related Works

We first review related works on discrimination against indigenous communities in Bangladesh to understand the reality. Due to the media's under-representation of these communities, the authors, residing in the capital city, had limited awareness of the actual conditions in the Chittagong Hill Tracts (CHT) and flatlands. Then we review on quantitative and qualitative methods used to identify discrimination of representation. Finally, we review how news articles are analyzed traditionally in literature.

### 2.1 Marginalization of indigenous communities in Bangladesh

Indigenous communities in Bangladesh, particularly in regions like the CHT and flatland areas, face significant challenges related to cultural preservation, human rights, and environmental sustainability. In 2011, United Nations Permanent Forum on Indigenous Issues (UNPFII) reported that the 1997 CHT Accord had hardly been implemented, noting de facto military rule and the excessive deployment of armed forces in the region. The UNPFII also recommended that the United Nations Department of Peace-keeping Operations (DPKO) prevent military personnel that are violating human rights from participating in UN peacekeeping missions. The report also mentions that human rights violations, committed primarily against the indigenous population, continue to be reported and seem to demonstrate a consistent pattern of human rights violations in the region. Violations include arbitrary arrests, torture, extrajudicial killings, harassment of rights activists and sexual harassment. In most cases such violations are carried out with impunity [68].

The issues of land grabbing are wide-spread. Seizure of private and common lands of the indigenous minority groups by powerful Bengali result in conversion of common lands into private property of Bengali settlers. Additionally, Bengali settlers use a range of different techniques for grabbing hilly-lands. In some cases, they forged land settlement documents to justify their (illegal) occupation. Otherwise, they simply occupied lands by force, given the backing of the security forces and the civil administration. Such incremental land grabbing by Bengali settlers has continued during the post-Accord period. Moreover, the common lands of the indigenous peoples in the USF (Unclassed State Forest) areas that they had traditionally used for Jhum cultivation (shifting cultivation), grazing, hunting and gathering, and other purposes are leased to members of the influential Bengali who do not even reside in the CHT. Indigenous people's customary rights over these lands were not recognized by the DC office, which treated these as state property and leased the lands to plantation leaseholders [41]. A total of 1,871 leases of 25-acre plots totaling 46,775 acres of "khas" (government owned) land in the CHT were issued prior to the Peace Accord, mostly to non-resident industrialists, companies, and civil and military officials, with only around 30 being granted to indigenous people [6, 43]. Conflict over lands is frequent and acute [17].

Indigenous people also face economic marginalization through indebtedness. In case of selling their produce, they have a double disadvantage as they do not have access to the bigger markets in Chittagong as these are controlled by the Bengali businessmen. In the large markets, they could get better prices for their produce but they have to depend on the Bengali middleman for trading there. Middlemen buy their produce and transport those to the large markets as the indigenous people themselves cannot afford to transport their produce to Chittagong. Therefore, they face economic loss and thus, become marginalized through indebtedness and land loss. Once they face economic loss, they fall in a spiral downward, either they are forced to sell land or to take loan at high interest rates, sometimes they take loan on the condition of advance selling of their produce (the price of advance sell is much lower than the actual price) [41].

The mainstream representation of indigenous culture often perpetuates stereotypes, using terms like "tribal" or "primitive," which reinforces existing hierarchies and undermines the complexity of indigenous identities [66].

Indigenous culture also faces threats. Zahed et al. reports that the settlers in CHT replaced the Chakma, Marma names of different places with Bengali and Muslim names. The author reported a list of such places [76]. Williams and Sowell discusses social ostracism, highlighting how indigenous minorities may experience psychological invisibility and exclusion, leading to marginalization and lack of recognition within society [74]. The marginalization of indigenous minorities stems from state repression, forced assimilation, and socio-economic changes threatening their culture and existence [4].

[26] found that sourcing of information for news coverage on Chakma (an indigenous community) migration to Tripura from Bangladesh was entirely based on state agenda through statements or interviews with government officials, ministers etc. There was absolutely no voice allowed to non state actors in the reports like victims of onslaught in the CHT, which later propelled the migration, or local residents of the area where the Chakma refugees had sought asylum in Tripura.

Violence against indigenous women (VAW) is one of the most widespread forms of human rights violation there. The VAW comes in many forms such as domestic violence, rape, assault, sexual harassment, prostitution, trafficking, abduction and forced marriage and early marriage of girls. Very few incidents were reported in leading national dailies and none of the accused received exemplary punishment. In most cases, the perpetrators of these gross human rights violations go unpunished even when the victims and witnesses identify them [13].

We can continue listing atrocities against indigenous people. Just [13] reports 32 VAW, 9 communal attacks, 1000+ injuries and even forced displacement from their homes, demolishing of 300+ indigenous houses and so on. Yet little to no affirmative actions are taken to address these issues.

## 2.2 Quantitative Analysis of News Article

Quantitative researches consider topic distribution of news articles, sentiment analysis to understand media representation. Additionally, keyword analysis, semantic role labeling, or dependency parsing can help determine the angle or perspective used to present information and thus aid in media framing analysis. Recent works use LLM to list topics of news to find topic distribution for a demography [57]. Earlier works use diverse approaches. Fonseca et al. investigated caste discrimination in India's context by employing a word2vec model trained on newspaper articles. They then constructed a correlation graph for the word "caste" based on cosine similarity scores. Instead of simply counting each occurrence of the word, they estimated the relative frequency of "caste" within each article by summing the cosine similarity scores of relevant words. Following this, they selected the top 7% of articles and applied Latent Dirichlet Allocation (LDA) to them. They concluded that there is an excessive association between lower castes, victimization, and social unrest in the news that does not adequately cover the reports on other aspects of their life and personal identity, thus reinforcing conflict [31]. Dursun uses grammatical analysis of language use in press to reveal different perspectives regarding the assignment of a more or less prominent role of the demography. The use of active and passive verb voice is one constituent of such analysis [29]. Yến-Khanh analyzed a 540,000-word corpus using WordSmith software to study language patterns. Their findings indicate that autism, a severely marginalized group, is predominantly framed as a medical issue and a family burden, rather than as a matter of social policy or a facet of human diversity [75]. Another study examined the discourse of discrimination composed by the media using topic modeling, focusing on news articles on discrimination reported in major national daily newspapers in Korea from 2010 to 2022 [51]. [21] employs a classifier to identify the political slants of news articles, followed by the use of word embeddings and the Linguistic Inquiry and Word Count (LIWC) tool to gain deeper insights into linguistic traits. Hayes et al. present a comprehensive approach, beginning with the collection of social media data on political discourse. They define what qualifies as political content and identify political tweets related to native communities. They then identify tweet frames of interest and visualize the distribution of these frames for both the general population and Native American communities, highlighting differences in the focus of political discourse. However, their approach relies heavily on network analysis methods [71]. Our quantitative methodology in Section 3 draws inspiration from this approach, where we use a similar approach for the identification of relevant articles, and frame analysis. While previous studies have computationally analyzed the under-representation of various demographics, such as gender, migrants, and people with disabilities, to the best of our knowledge, no such research has been conducted specifically for the indigenous communities of Bangladesh.

## 2.3 Qualitative Analysis of Media Representation

Discrimination of representation in media has been widely studied qualitatively. We review some of these works and found that interviews, controlled social experiments and surveys have been used for qualitative research. Galdi et al. showed through social experiments how stereotypical gay men representation increases social bias about them. Heterosexual Italian men (N = 158) were exposed to a clip portraying (i) a stereotypical feminine gay male character, (ii) a counter-stereotypical masculine gay male character, or (iii) a nature documentary. Compared to the other conditions, exposure to the counter-stereotypical gay character increased discrimination toward gay men, in the form of anti-gay jokes, the higher the level of participants' prejudice against gay men. Results further demonstrated that this effect was explained by reduced perceived stereo-typicality of the character [33]. Saleem and Ramasubramanian found that Muslim American students who viewed negative media representations of their religious ingroup, relative to a control video, were less likely to desire acceptance by other Americans and more likely to avoid interactions with majority members. Such discrimination of representation increase inter-group distance [63]. Haraldsson showcased the way media discrimination hinders progress towards putting femininity on an equal footing with masculinity in the political domain in her thesis [40]. Ittefaq et al. revealed that sanitary workers in Pakistan believe that they do not have any representation in Pakistan's mainstream media to voice their issues. Moreover, they have serious reservations about their polemic social representation and voice concerns regarding the media that often amplify such depictions [42]. Balasubramaniam et al. shows how discrimination of representation affects the Dalit community. They argue that under-representation of Dalits leads to news exclusion related to dalits [10]. Atuel et al. leveraged topic distribution, article count and article length as a means to understand media representation [8].

### 2.4 Media influence on people

Media not only reflects societal values but also actively molds them. In our analysis, we apply key theoretical frameworks such as news framing, priming, mobilizing and agenda-setting to understand the extent of media influence. These concepts are foundational in media studies and communication research, having been extensively studied and supported by scholars across the field. They are essential to understanding how news media portrays a news about a demography.

McCombs and Shaw describe agenda setting as the media's ability to influence the salience of topics in the public mind. By focusing on specific issues, the media determines which topics become the focus of public attention, shaping the public's perception of what is important. This process does not dictate opinions but rather guides which issues people consider most pressing in society [52]. For example, if the media consistently highlights climate change as a top issue by covering it frequently, running stories about its impacts, and featuring it in headlines, the public begins to perceive climate change as one of the most pressing issues. Even if individuals do not know the details, they start considering climate change a priority because it is continually emphasized by the media.

According to Entman, framing is the process of selecting certain aspects of a perceived reality and making them more salient in a communication text. This involves promoting a particular problem definition, causal interpretation, moral evaluation, and/or treatment recommendation. Media frames influence how people understand and interpret issues by shaping the context and meaning of the message [30]. Iyengar explain priming as the process by which media exposure affects the standards by which people evaluate issues or political actors [44]. Repeated coverage of certain issues leads audiences to use those issues as a benchmark for evaluating political performance or other related subjects. Essentially, media primes the audience to make judgments based on the specific topics it emphasizes [45]. For example, after climate change becomes a prominent issue in the media (agenda setting), priming affects how people evaluate political leaders or policies. For example, during an election, voters might prioritize a candidate's stance on climate change over other topics like the economy or healthcare, because the media has primed them to use climate change as a key factor in their judgment. Norris defines mobilizing as the ability of media to inspire people to take action, often in a political or social context. Through reporting and highlighting certain events, issues, or injustices, the media can spur collective action, such as protests, voting, or advocacy efforts, by raising awareness and encouraging participation [54].

But how can we analyze them in news articles? Analyzing agenda setting is straight-forward, since the bias can be detected through counting the occurrences of different stories/themes within the news articles [72]. Similarly, [35] uses distribution of key themes to understand agenda setting. However, these methods are not applicable to biases which are implicit in wording, namely, "framing" bias. Automatic detection of framing bias is challenging since nuances in the wording can change the interpretation of the story. Morstatter et al. first curates a labeled dataset on framing containing 823 news articles. We note that the frames were decided by the annotators based on the corpus. They then used logistic regression, n-gram, LSTM classifiers to predict frame given a news article [53]. Another work from Pew Research Center also identified frames based on the corpus [32]. Such an approach is specific to the demography being studied. Recent works also use LLM for news frame identification [27]. [37] was able to show a connection between lexical semantics and readers' perception. This is important as they proposed a computational approach to identify this relationship based upon sentiment. As for news priming, [5] proposed a cognitive model to identify priming in a news article. They first collect priming effect against different dosage of stereotypes based on the cognitive model. Then they fit different models in an attempt to understand the correlation between priming and dosage factors. Similarly, mobilizing also relies on mostly manual news content analysis [77]. Recent works on LLM for news discourse analysis worked on identifying actors in a discourse network but it did not focus on identifying claims. Some claims are mobilizing statements i.e. inspiring people to take action [11]. We consider mobilizing out of scope in this study.

## 3 Methodology

We use both quantitative (Section 3.1) and qualitative (Section 3.2) methods to understand the presence and impact of discrimination in representation on indigenous people of Bangladesh.

### 3.1 Quantitative Method

We use various Natural Language Processing (NLP) techniques to find quantitative support for discrimination of representation. Our source of data is the ebD dataset, containing 2.2 million news articles from 3 of the most reliable/popular online and publicly available newspapers in Bangladesh [15]. We first identify indigenous articles from the dataset using GPT-3.5 Turbo and keywords. We then estimate news genre distribution using CorEx [34], a semi-supervised topic model. We use news genres identified Table-2 as anchors. We also estimate news sentiment and average article length to get more data on news representation. Finally, we use GPT-4o-mini to estimate news framing and word cloud to approximate news priming. We also try to explain agenda setting based on our news genre distribution estimated before.

#### 3.1.1 Identifying articles on indigenous people

To identify articles on indigenous people, we first need to define what is considered an article on indigenous people (here-forth called *Indigenous article*). To the best of our knowledge, no such definition exists in literature. And since it is a subjective concept, there can be various possible definitions. We take inspiration from [65] and [56] on considering a narrow definition and a broad definition in situations where many definitions can exist.

3.1.1.1 Broad definition

The broadest definition of indigenous article can be - any mention of indigenous people in an article. Because an article is unlikely to be about indigenous people if it does not mention them explicitly, either syntactically or semantically. [31] uses word embedding to find semantic references to a demographic group within an article. Due to the sheer volume of articles (2.2M), we instead use a syntactic method, specifically keyword-based filtering.

We first compiled a list of indigenous communities in Bangladesh from Wikipedia [3]. However, we later excluded certain tribe names, such as "বম" (Bawm), "কোচ" (Koch), and "খাসি" (Khasi) from our keyword list because they also function as prefixes or suffixes in other Bangla words (e.g., "খেলার কোচ" (game coach)).

Table 1. Evaluating performance of LLM in indigenous Article Classification.

| Model | Prompt | Accuracy | Precision | Recall | F1 score |
|---|---|---|---|---|---|
| GPT-3.5-turbo, ctx10k | ethnic-only-prompt.txt | 62% | 71.9 | 65.1 | 68.3 |
| GPT-4o-mini, ctx 10k | ethnic-only-prompt.txt | 78% | 81.5 | 84.12 | 82.8 |
| GPT-3.5-turbo, ctx10k | combined and formal prompt.txt | 72% | 72.15 | 90.5 | 80.2 |
| GPT-4o-mini | combined and formal prompt.txt | 67% | 65.5 | 100 | 79.24% |
| GPT-4o-mini, ctx 10k | combined and formal prompt.txt | 69% | 67.02 | 100 | 80.25 |
| GPT-4o-mini, ctx 10k | combined and informal prompt.txt | 70% | 67.7 | 100 | 80.7 |
| GPT-3.5-turbo, ctx10k | combined and informal prompt.txt | 72% | 76.9 | 79.4 | 78.13 |
| GPT-3.5-turbo, ctx10k | combined and indigenous.txt | 77% | **81.3** | **82.5** | **81.9** |

We initially identified 10187 articles containing these keywords. We then observed that a significant number of articles are on the Rohingya crisis [55]. This indigenous community is not from Bangladesh so they are out of scope for this study. So we removed articles on Rohingya crisis using keyword-based filtering. After further refinement, we were left with 4,893 articles that included our selected keywords. We refer to this collection as *KeywordDataset*.

3.1.1.2 Narrow definition

A narrow definition for indigenous article can be - Any article that focuses entirely on indigenous people. We use the emergent capabilities of large language models (LLMs) to classify indigenous articles based on a narrow definition. However, We first need to validate that the large language model (LLM) can achieve satisfactory performance.

We started off by randomly sampling 100 articles. Two annotators were instructed to label each article as True or False, representing whether the article is on indigenous people or not. We found 75% agreement rate between them. After manual inspection, we discovered that the disagreement between annotators was on political articles. The annotators disagreed on whether an article on political news is an indigenous article or not. After discussion, the disagreement was resolved. Any political article with indigenous people's involvement represents the political aspect of their life. After the annotation process, we obtained 63 "True" labels and 37 "False" labels from the 100 articles.

We then experimented with different prompts with GPT-3.5 Turbo and GPT-4o mini. We observed significant difference in performance when we used different prompts. The *indigenous-only-prompt* is a prompt to classify a news as being an indigenous article or not. The other prompts combine news description style, news sentiment, news agenda setting, framing, priming, mobilizing with indigenous article classification. We combine them into one prompt/request because annotation using ChatGPT is costly. We experimented with various prompts to enhance performance. The details of these prompts can be found in the supplementary document. Our prompt with informal language i.e. *combined and informal prompt.txt* achieved moderate performance while a more formal, well-defined prompt (*combined and formal prompt.txt*) resulted in slight improvement in F1-Score (Table 1). In Bangladesh, the terms "ethnic minority" and "indigenous" are used synonymously [61]. So, we considered replacing the word "ethnic minority" with "indigenous" and observed significant performance improvement in precision, recall and F1 score. We believe this is due to the term 'indigenous' being more commonly used in English compared to 'ethnic minority' which may result in large language models (LLMs) having a better understanding of the word 'indigenous'. Given that the GPT-3.5 Turbo model has a context window length of 16385, a significant portion of this is consumed by our prompt itself. So we consider 10,000 context length of each article. This is also to limit the associated cost of labeling. Our experimentation results are summarized in Table 1. Note that, ctx means context length.

We then began annotating news articles based on a narrow definition of indigenous articles using the best-performing model, GPT-3.5-turbo, along with its associated prompt. We consider 4893 articles from KeywordDataset (Section 3.1.1) for annotation and 5K articles from the ebD corpus via random sampling. The results of annotation is discussed in Section 4.

### 3.1.2 News Genre Distribution

News genre/topic distribution has been widely used in literature to understand discrimination of representation[31, 51, 57]. We first identify major news genres and then computationally estimate the representation in each news genre.

#### 3.1.2.1 Identifying News Genre

AP News Taxonomy is a classification system for English-language news content. Their classifications are globally recognized and utilized by numerous media outlets [7]. However, some of their classes, including U.S. news, World news, Washington news do not align with Bangladesh. We also found no evidence from literature review that Bangladeshi newspapers follow any such standards. So we decided to identify news genre by analyzing which genres Bangladeshi newspapers mostly focus on. For that, we first identify the news genres used by the three most popular newspapers in Bangladesh by browsing their respective websites. We then pick news genres through majority voting from them, as we can see in Table 2.

To ensure openness to emerging news themes and genres, we experimented for several rounds and manually examined the articles under each topic. After analysis, we later added 2 more genres to the list, namely "Protests and Social Movements", and "Govt Actions". The reason is - these genres are fundamental to understanding the representation of indigenous people in news media.

#### 3.1.2.2 Estimating News Genre Distribution

We use semi-supervised topic modeling to estimate News Genre Distribution. Because after several rounds of experiments, we observed that unsupervised topic modeling can sometimes generate themes that are not meaningful or directly relevant. So we used CorEx topic modeling [34] on 4893 indigenous KeywordDataset and achieved a topic coherence score of 143.63. We also performed CorEx topic modeling on 100K randomly sampled Bangla news articled and achieved a topic coherence score of 55.49. This lower score is due to the diversity of topics in 100k Bangla news articles. This further shows support for our hypothesis of discrimination of representation. Our results of topic modeling is discussed in Section 4.

### 3.1.3 Estimating News Sentiment

We evaluated GPT-3.5-turbo performance on news sentiment analysis. We annotated 109 randomly sampled news articles through two human annotators. The agreement score was 86.24%. GPT-3.5-turbo achieved an accuracy and F1 score of 74%, and 77% respectively. We then used GPT-4.5-turbo to annotate 9893 news articles, the description of this curation process is in Section 3.1.1.

### 3.1.4 Estimating Media Influence

For the longest time, newspaper articles have mostly been analyzed manually by researchers. This has restricted the size of data they can process. So in this work, we use LLM to analyze news articles in an attempt to scale the processing, particularly focusing on the impact that news articles have on public perception. As discussed in Section 2.4, we consider agenda setting, news framing and priming to understand media influence on people.

Table 2. Identifying Bangla News Genre

| Prothom Alo | The Daily Star | Bangladesh Protidin | News Genre |
|---|---|---|---|
| Politics (রাজনীতি) | Bangladesh | Politics (রাজনীতি), National (জাতীয়) | Politics |
| Crime (অপরাধ) | Investigative story | - | Crime |
| International (বিশ্ব) | World, Asia | East-West (পূর্ব-পশ্চিম) | International Affairs |
| Business (বাণিজ্য) | Business | Business (বাণিজ্য) | Business & Economy |
| Sports (খেলা) | Sports | Field (মাঠে ময়দানে) | Sports |
| Entertainment (বিনোদন) | Entertainment | Showbies (শোবিজ) | Culture & Entertainment |
| Jobs (চাকরি) | Youth ->Career | | Jobs |
| Lifestyle (জীবনযাপন) | Life&Living ->Fashion, Food, Family, Travel | City Life (নগর জীবন) | Lifestyle |
| Bangladesh (বাংলাদেশ) ->Corona (করোনাভাইরাস) | Life&Living ->Health & Fitness | Health Corner (হেলথ কর্নার) | Health |
| Bangladesh (বাংলাদেশ) ->Environment (পরিবেশ) | Environment | - | Environment |
| Bangladesh (বাংলাদেশ) ->Main city (রাজধানী), state (জেলা) | Bangladesh | Local news (দেশগ্রাম), (Daily Chittagong) চট্টগ্রাম প্রতিদিন | Local news |
| Education (শিক্ষা) | Youth ->Education | Campus (ক্যাম্পাস) | Education |
| Technology (প্রযুক্তি), Gadget (গ্যাজেট) | Tech&Startup ->Gadgets, Gaming, Guides, Startups | Tech World (টেক ওয়ার্ল্ড) | Technology |
| Religion (ধর্ম) | - | Islamic Life (ইসলামী জীবন) | Religion |
| (Science Thoughts) বিজ্ঞানচিন্তা | - | Science (বিজ্ঞান) | Science |

**Agenda Setting:** Agenda setting is the media's ability to influence the salience of topics in the public mind. If media highlights certain topics consistently, that is likely the agenda setting. So we try to infer it from news genre (frequency) distribution, as discussed in Section 2. We note that literature identifies agenda themes through domain knowledge and key topics in society in a time-frame [35, 72]. Due to the sheer volume and time-frame of the eBD news articles, we opt to use more generic themes identified in Section 3.1.2.

**Framing:** Framing are biases which are implicit in wording. It is analyzed by focusing on wording, tone of news articles. We employ two approaches from the literature to identify frames in each article. One approach directly utilizes frames previously reported in the literature, while the other focuses on extracting frames through the analysis of articles. The first approach uses the 5-frame classification proposed by Barrio et al. [27]. Following their methodology, we have considered the following five frames.

- Responsibility: Its commonly known as *Attribution of Responsibility*. This frame presents a problem or issue in such a way as to attribute responsibility for its cause or solution to either the government or to an individual or group.
- Human: Its commonly known as *Human Interest*. This frame brings a human face or an emotional angle to the presentation of an event, problem, or issue.

- Conflict: This frame emphasizes conflict between individuals, groups, or institutions as a means of capturing audience interest.
- Morality: This frame puts the event, issue, or problem in the context of religious tenets or moral prescriptions.
- Economic: This frame reports an event, issue, or problem in terms of the consequences it will have economically on an individual, group, institution, region, or country.

We utilize GPT-4o-mini to categorize each article according to the five frames mentioned above. We have used the same prompt as [27] and achieved a human-machine agreement rate of 58% on a validation set of 109 hand-annotated samples. Due to the complexity and objectivity of identifying news frames and limited funding, we opted to use a single annotator similar to [27]. The annotator is well-versed in historical backgrounds and indigenous struggles. Although they received an extensive briefing on how the frames should be attributed, the results are nevertheless influenced by personal preferences and experiences. Our 58% agreement rate represents an improvement over the 43% reported in [27], likely due to our use of GPT-4o-mini instead of GPT-3.5.

Secondly, following the methodologies outlined in [53] and [32], we first analyze the articles to identify frames. We use GPT-3.5-turbo to independently analyze the frames within the news articles, resulting in the identification of 1,626 frames. To focus on the most prevalent frames, we select those that appear in at least 10 articles. From the initial set of 7,787 annotated articles (excluding those with missing values), 6,174 remained after applying this threshold, leading to the identification of 68 frames. A manual review further consolidated these into 16 significant frames. More specifically, We observed that some frames are inter-related so we iteratively combined them. We then validate the annotation of these 16 frames using a sample of 109 news articles. According to [53], framing is inherently nuanced. The intertwined nature of frames often leads to confusion among annotators, as a single news piece can contain multiple frames [27]. This complexity makes it challenging to ensure that all frames are accurately captured during annotation. Additionally, a machine may detect a subset of frames or identify an overlapping but distinct set of frames. This scenario is best treated as a fuzzy or soft evaluation problem. A "soft match" occurs when a machine's prediction is considered correct if it matches any of the multiple frames identified by human annotators, even if the match is not exact or comprehensive. The Jaccard Index is used to estimate similarity between sets [64], making it suitable for our soft match evaluation. Our analysis yielded a human-machine agreement score of 0.37 using the Jaccard Index on 109 hand-annotated news articles. Given the low level of agreement, we have opted not to include the detailed findings from our annotation of frames across 9,893 articles in our final analysis.

**Priming:** Priming is the process by which media exposure affects the standards by which people evaluate issues or political actors. Repeated coverage of certain issues leads audiences to use those issues as a benchmark for evaluating political performance or other related subjects. There are no previous works that automatically detects priming in a news article to the best of our knowledge. So based on priming definition, we first take political articles (using CorEx) and create word cloud from them to understand which issues are repeatedly covered about political actors. Note that, both protests and politics genre contain political aspect of indigenous people according to participants in Focus Group Discussion. More details on it in Section 3.2.

### 3.1.5 News Article Length

Atuel et al. used topic distribution, indigenous article count and indigenous article size to understand media representation [8]. So we also hypothesize that indigenous news articles are shorter in length than general news articles. We test this hypothesis on 4893 news articles each, using Welch's t-test [28] with a level of significance of 5% (p=0.05). The result is summarized in Section 4.3.

#### 3.1.6 Summary of Quantitative Method

To summarize, our approach begins with using LLMs to identify articles related to indigenous communities based on both narrow and broad definitions. Following this, we apply CorEx semi-supervised topic modeling to uncover the news genre distribution of indigenous and general news articles. Note that, We identified the news genres through manual analysis of three prominent Bangla newspapers. We also perform sentiment analysis on articles concerning indigenous communities to assess their tone. To evaluate the influence of media, we analyze agenda-setting (leveraging the previously identified news genre distribution), framing (using LLM), and priming (through word cloud analysis). The results of these analyses are detailed in Section 4. Next, we present the qualitative methods employed in this study.

### 3.2 Qualitative Method

We use Focus Group Discussion (FGD) as a qualitative method to understand discrimination of representation of indigenous people and its impacts. This method both validates our findings from quantitative method and uncovers the consequences of discrimination of representation.

#### 3.2.1 Question Set

We aimed to keep the focused group discussion as open as possible in an attempt to get more insights into real situation. So we kept our questions to a minimum of 6. We also kept the questions in second person to make the participants feel more engaged.

- Do you feel well-represented in news?
- Do you feel that critical news about your community are not highlighted in media?
- Do you feel there is a barrier that filters your news?
- Do you feel that media helps perpetuate stereotypes about you?
- Any other experiences you want to share?
- What can news media do to address such discrimination?

We elaborated these questions further in a question set available to authors during FGD. One of the authors is from indigenous community and they validated the question set to remove any discriminatory terms. The final question set has a total of 24 questions, included in Appendix-A. We, however, did not strongly follow through all questions to keep the conversation natural and let the participants speak their heart on key concerns.

#### 3.2.2 Recruitment

Majority of indigenous communities live in hilly regions in Bangladesh i.e. Rangamati, Khagrachari and Bandarban. Additionally, a part of the population also lives in flatland i.e. Sylhet, Mymansingh, Rajshahi [24]. These regions are far away from the capital. Additionally, focusing on a particular region would result in not understanding the whole population. So we conducted a virtual FGD. In this way, we easily connected with participants from all the hilly regions and some flatland regions. Moreover, the virtual setting also addressed safety concerns. With newfound independence after a mass uprising of Gen-Z, current situations in Bangladesh had yet to stabilize [22]. So a virtual discussion was a safer option.

The indigenous author in our research group used snowballing to identify 15 contact persons of indigenous community, among which 9 gave their consent to participate.

### 3.2.3 Participant Background

We collected participants' demographic information anonymously through Google Forms, after their informed consent. The participants come from Chakma, Garo, Khumi and Marma indigenous groups in Bangladesh. 33% participants are indigenous women and they are in 20-22 age range. We had participants from both hill tracts and flatland. 77.8% reported to consume news media on a regular basis. 77.8% participants consume online or print media news. While the rest relies on Facebook news.

### 3.2.4 Data Analysis

Two hours of FGD meeting recording was documented into a meeting moments format manually. All perspectives raised by participants were covered in the meeting moments. The meeting moments document was also made de-identifiable against participants since no information pertaining to personal information was documented in the meeting moments.

The meeting moments document was validated by the indigenous author of the research paper. The document was also shared with the participants to ensure that we highlighted the discussions in the same perspective they presented.

### 3.2.5 Ethical Considerations & Positional Statements

Throughout the discussions, we were extremely careful to have a respectful conversation with our participants as they belong to a highly marginalized community with their own culture and norms. Before starting the FGD, we engaged in a long conversation with a member of indigenous community. We learned of current situations to better derive questions for the FGD. We also got our questionnaire reviewed by the indigenous community member.

## 4 Findings

In this section, we present a detailed analysis of our findings obtained from the quantitative and qualitative methods outlined in the previous section.

### 4.1 Strong support of Discrimination of Representation

The news genre distribution on 4893 indigenous KeywordDataset and 100K randomly sampled Bangla news article is shown in Figure 1. We observe a significant discrimination of representation between two groups. Specifically, there is a focus on culture & entertainment, crime and protests in indigenous news. We validated these findings in three ways. Firstly, we observe that our general news genre distribution derived using semi-supervised CorEx topic modeling matches with news genre distribution of existing literature. Table 3 shows news genre distribution by [59]. As we can see, Social and legal issues are the highest, followed by politics and government. This is also the same in our Figure 1 for general news distribution in Bangla. Secondly, we asked the participants which genre of news about the indigenous people do they encounter the most in newspapers. Table 4 presents their responses. The most common responses were "Culture and Entertainment", "Lifestyle", "Protest and Movement", "Govt action". This is very similar to our findings in Figure 1. Thirdly, in our survey, all participants unanimously agreed that they feel strong under-representation and mis-representation in media. As we can see, we find strong support for discrimination of representation of indigenous people in Bangladesh news media.

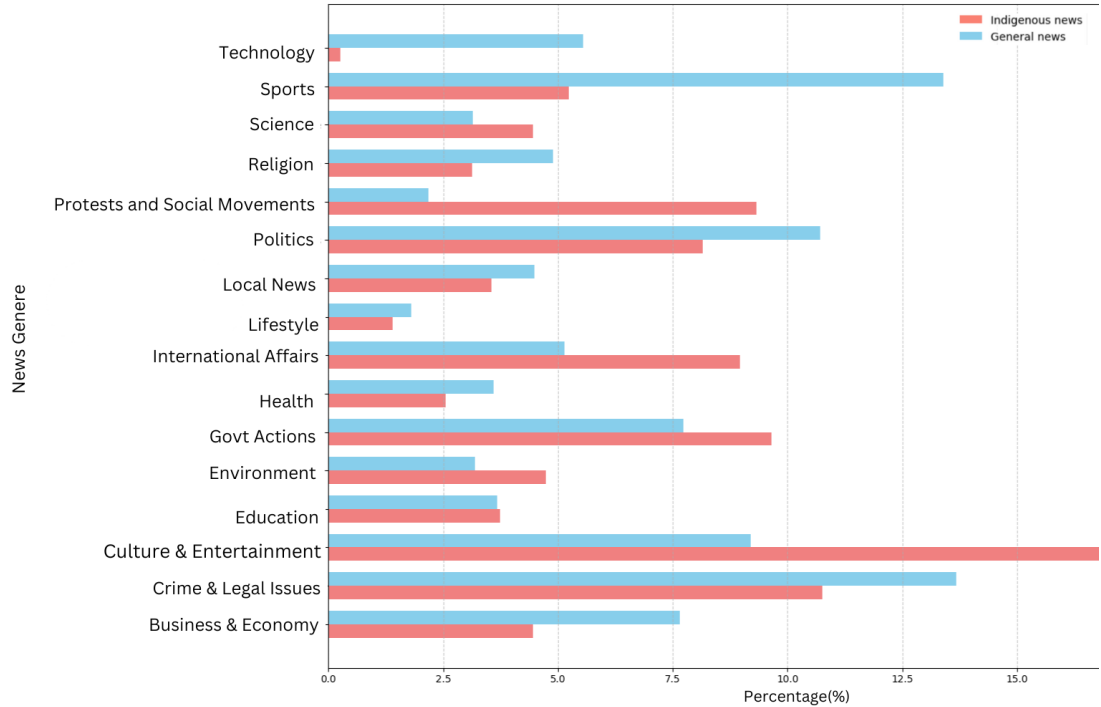

Fig. 1. Genre Distribution in Bangla News: Indigenous vs. General Focus

Table 3. News Genre Distribution in Literature [59]

| Summary news category | Distribution |
|---|---|
| Politics and Government | 23% |
| Economy | 11% |
| Science and Health | 11% |
| Social and Legal | **32%** |
| Crime and Violence | 10% |
| Celebrity, Arts, Media, Sport | 13% |
| Other | 0 |

Table 4. Perceived news genre distribution by indigenous people

| News Genre | Perceived Representation |
|---|---|
| Culture & Entertainment | 88.9% (8 responses) |
| Lifestyle | 33.3% (3 responses) |
| Business & Economy | 11.1% (1 response) |
| Local news | 11.1% (1 response) |
| Protests & Social Movements | 22.2% (2 responses) |
| Govt. Actions | 22.2% (2 responses) |
| Crime | 11.1% (1 response) |
| Politics | 11.1% (1 response) |

### 4.2 Prevalence of Negative Sentiment

We also note a significant prevalence of negative sentiment in the indigenous articles annotated by the GPT-3.5-turbo, as illustrated in Figure 2. Specifically, 57% of news articles about indigenous people convey negative sentiment, compared to only 28% for general news. While nearly 49% of general news articles are neutral, only about 7.5% of indigenous articles exhibit neutral sentiment. Our FGD findings also indicated that participants feel the news media portray their stories in a predominantly negative light. This observation signifies that indigenous narratives in the media are predominantly framed in a negative light, which could reflect broader societal biases or stereotypes. The higher percentage of negative sentiment suggests that coverage of indigenous issues may often focus on problems, conflicts, or crises rather than positive stories or achievements. The low neutral sentiment in indigenous articles compared to general news indicates a lack of balanced representation, potentially contributing to a skewed public perception of indigenous communities.

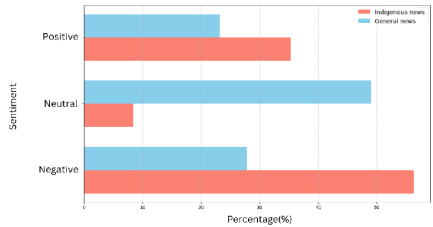

Fig. 2. Sentiment comparison between indigenous news and general news

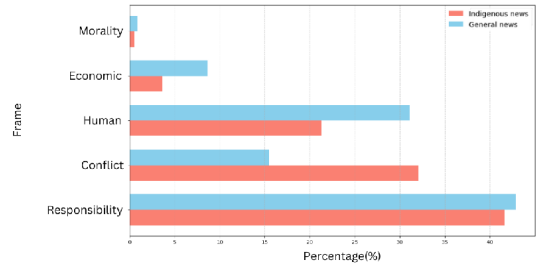

Fig. 3. Comparison of news framing focus (5 frames from literature)

### 4.3 Assessing Article Length in Indigenous News

We considered mean length of 4893 indigenous and randomly sampled 4893 general news articles. Our hypothesis testing (Welch's t-test, p-value = 0.05) found no support for the hypothesis that indigenous articles are smaller than general articles. This finding is somewhat contrary to literature where [8] reports smaller article length for black people compared to white people.

### 4.4 Shaping Perceptions of Violence about Indigenous Communities

The concepts of news framing, priming, and agenda setting are foundational in the fields of media studies and communication. We used LLM to automate identification of framing, priming and agenda setting. The findings are discussed below.

#### 4.4.1 Agenda Setting

The key agendas in indigenous articles can be inferred from the news genre distribution shown in Figure 1. Notably, Culture & Entertainment and Protests & Social Movements emerge as the primary agendas. The focus on Culture & Entertainment highlights the media's portrayal of indigenous cultural expressions and traditions, promoting awareness but also risking commodification. In contrast, the prominence of Protests & Social Movements underscores the sociopolitical struggles faced by indigenous communities, such as land rights and social justice issues

#### 4.4.2 Framing

We automate frame analysis using LLM in Section 3.1.4. Our findings from 9893 articles are in Figure 3. As we can see, the conflict frame is predominant in the representation of indigenous news articles, whereas the human interest frame is more common in general news. This suggests that indigenous news articles are more often portrayed through a lens that emphasizes tensions, disagreements, or struggles, while general news tends to focus more on stories that highlight individual experiences, emotions, or personal stories. This difference in framing could reflect underlying biases in how media outlets approach different types of stories, potentially influencing public perception of indigenous communities and their issues.

### 4.4.3 Priming

The results of CorEx in 2873 indigenous news articles and then creating word cloud for protests and political topic is shown in Figure 4a, and 4b.

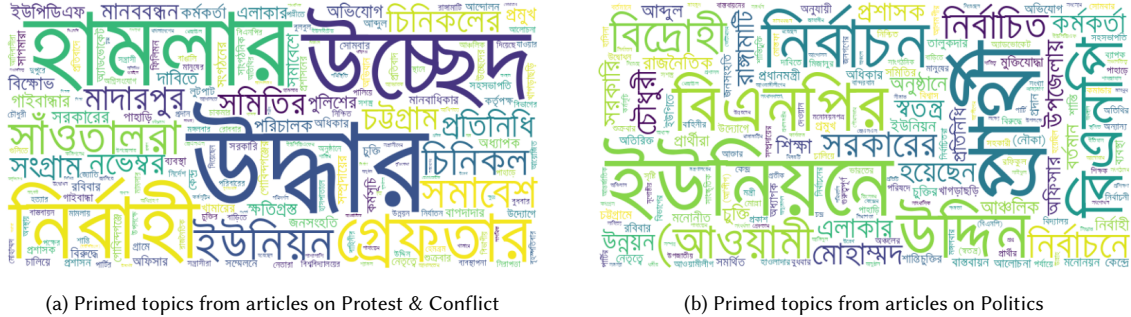

(a) Primed topics from articles on Protest & Conflict

(b) Primed topics from articles on Politics

Fig. 4. Priming Analysis of Different News Genre

The observation is that in political articles, words like উচ্ছেদ (eviction), উদ্ধার (rescue), হামলা (attack), গ্রেফতার (arrest), চিনিকল (sugar mill), নির্বাহী (executive), ইউপিডিএফ (UPDF - United People's Democratic Front), নির্বাচন (election), প্রার্থী (candidate), and বিদ্রোহী (rebel) are frequently highlighted in a word cloud indicates a focus on political instability, law enforcement, economic issues, and election dynamics. The terms উচ্ছেদ, উদ্ধার, হামলা, গ্রেফতার, and বিদ্রোহী point to issues related to social unrest, conflicts, and government intervention. Frequent use of these words in articles could prime readers to associate political scenarios with security concerns and civil disturbances. The inclusion of চিনিকল (sugar mill) suggests that industrial issues or labor disputes may be a part of political discussions. This could prime attention toward economic challenges or governmental roles in managing industries. Words like ইউপিডিএফ, নির্বাচন, প্রার্থী, and নির্বাহী indicate a significant focus on elections, political candidates, and political parties. The priming effect here may lead readers to focus more on political competition, leadership, and governance.

## 4.5 Critical news are not highlighted

In the FGD, participants discussed a total of 13 issues that are critical to them but they feel they are not properly highlighted in news media. Among them, we pick 5 issues that were discussed by multiple participants to reduce bias.

### 4.5.1 Discrimination towards Women

News articles such as "চাকমা মেয়েরা কেন এত সুন্দরী হয়?" ("Why are Chakma women so pretty?") highlight the selective focus of media on indigenous communities in a superficial or stereotypical manner. This kind of coverage emphasizes exoticism and beauty stereotypes, framing indigenous groups in ways that may seem positive but ultimately perpetuate reductive and often harmful images of marginalized communities. Participants observed,

> News like "Why are Chakma women so pretty?" are covered very religiously in almost all news media.
> But news on sexual assaults, oppression are rarely covered.

On the other hand, critical issues such as sexual assaults, oppression, and systematic discrimination often go underreported or are framed less prominently in the news. This gap in coverage can be understood through agenda setting theory, which suggests that the media dictates public priorities by focusing on certain issues over others. Another participant also reported their experience.

> During the flood, three girls were raped, including a minor. I could see no report of it in any news. So I approached a news media with conclusive proofs of the incident. But the news media said that it is not a relevant news.

The under-reporting of sexual assaults and oppression, particularly in indigenous communities, reflects a structural bias in the media. Galtung and Ruge's theory of news values can be applied here, as it explains how certain stories are deemed more newsworthy based on factors such as prominence, conflict, and cultural proximity [47]. When it comes to indigenous communities, issues that challenge societal norms, like oppression or sexual assault, may be deemed less newsworthy compared to stories that fit into pre-existing stereotypes. Studies on media bias in indigenous representation have shown that stories which challenge the status quo or criticize institutional oppression are less likely to receive proper coverage [16]. This often results in the silencing of marginalized voices, limiting the ability of the public to engage with the true challenges facing these communities.

#### 4.5.2 Invisibility during Flood

Participants frequently pointed out instances where local indigenous communities were overlooked by news media during significant events, such as natural disasters. This sentiment was especially relevant given the recent flood that hit Bangladesh just weeks before the FGD. On August 21, 2024, heavy rainfall, combined with a surge of water released from a dam in India's Tripura, triggered severe flooding across 73 upazilas (sub-districts) and 528 unions/municipalities in 11 districts in northeastern and southeastern Bangladesh. The National Disaster Response Coordination Center (NDRCC) reported that around 5.8 million people were affected, making it one of the most severe flood events in recent history [73]. A specific example shared during the FGD was a news report on the flooding in Sajek—a scenic valley in the Chittagong Hill Tracts (CHT)—which focused predominantly on the difficulties faced by stranded tourists (Figure 5).

The news on "সাজেকে ডুবে গেল সড়ক , আটকা পড়েছেন শতাধিক পর্যটক" ("The road in Sajek is flooded, over a hundred tourists stranded") detailed the inconvenience faced by tourists but made no mention of the indigenous people residing in the area, many of whom suffered far worse due to the flood. Participants expressed frustration that while the discomfort of tourists was highlighted, the severe struggles of the local population were ignored. This omission was contrasted with media coverage of similar crises affecting non-indigenous communities. For instance, in a different report covering floods in Feni, the headline read "ফেনীতে বন্যায় শতাধিক গ্রাম প্লাবিত, লক্ষাধিক মানুষের দুর্ভোগ" ("Hundreds of villages flooded in Feni, hundreds of thousands of people in distress"), placing the focus on the widespread human suffering. The participants pointed out that when indigenous people are involved, the narrative often shifts to focus on external or majority group concerns, leaving the indigenous communities and their challenges invisible to the broader public.

This pattern of exclusion in news reporting reflects what Gaye Tuchman refers to as symbolic annihilation [67], where certain groups are rendered invisible or marginalized in media narratives. When crises are reported, media focus tends to prioritize the experiences of tourists or majority populations, neglecting the unique struggles of indigenous minorities. Such selective reporting was described by the participants as a form of "media erasure," reinforcing

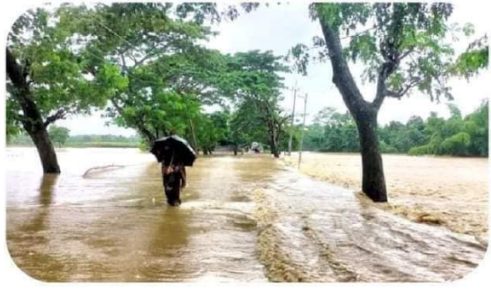

(a) Indigenous News (Translation: "Roads in Sajek have submerged, leaving over a hundred **tourists** stranded.")

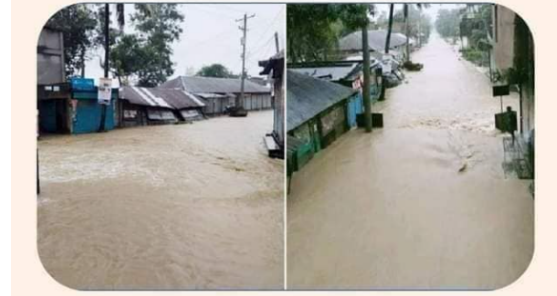

(b) General News (Translation: "In Feni, over a hundred villages have been flooded, causing suffering for hundreds of thousands of people.")

Fig. 5. Lack of Visibility of Indigenous Communities in Flood Coverage

existing power structures and perpetuating systemic inequalities [70]. These findings echo broader scholarship on the misrepresentation and under-representation of indigenous minorities in news media. For example, Cottle et al. argues that news media often focus on the most visible or sensational aspects of crises, further marginalizing less prominent, though equally affected, communities [25]. Similarly, other scholars have observed that media coverage of disasters often emphasizes the impact on tourists and urban centers while neglecting rural and indigenous populations [14].

#### 4.5.3 Human Rights Violation

The indigenous people are subjected to severe restrictions that infringe on their basic human rights. These practices create an environment of constant surveillance and control, making daily life challenging for the indigenous communities. Participants said,

> Indigenous people are required to obtain the signature of local authorities before they can purchase anything. Moreover, they are limited to buying no more than five kilograms of rice in a single transaction. This restriction poses a significant hardship, as five kilograms of rice is insufficient to last even a week, especially for larger families. The mountainous terrain makes it difficult to make frequent trips to markets, exacerbating the inconvenience.

This highlights how the daily economic activities of the indigenous people are controlled, severely limiting their autonomy. The need for approval for routine purchases suggests an oppressive environment, where even basic necessities are not easily accessible. The restriction on the quantity of rice further undermines their ability to secure sufficient food, placing an unfair burden on them due to geographic and economic factors. Participants also mentioned,

> To enter our homes (when traveling to and from the CHT), we are required to show an ID card and have a video recording taken of our face. Even if a woman is wearing a mask and is uncomfortable revealing her face, she is compelled to remove the mask and show her face for the camera.

[20] also reports similar experiences of being subjected to 'physical checks.' This practice of ID checks and mandatory video recordings reflects an extreme level of surveillance. Forcing individuals, particularly women, to reveal their faces — even when they are uncomfortable — violates personal privacy and autonomy. It also highlights gendered implications, as women may face heightened discomfort or cultural restrictions regarding the exposure of their faces in public. This creates a climate of constant monitoring and control, depriving the indigenous population of their right to privacy within their own homes and communities. The participants expressed that distressing events affecting their communities, such as this, often go unreported in the media, highlighting the gross discrimination of representation.

#### 4.5.4 Land grabbing

Land grabbing, particularly in indigenous communities, is a pervasive issue where powerful individuals or entities forcefully acquire land, often in the name of development or tourism. As a participant said,

> Indigenous Lusai communities are subjected to forced displacement as their ancestral lands are reclassified as "reserved" for tourism and development. This is just one example of a broader pattern of land grabbing by local powerful individuals.

We have also identified similar instances of land grabbing in International Land Coalition movement [23]. For example, the Mro community in Bandarban has faced land grabs linked to the construction of a five-star hotel, threatening their ancestral lands and livelihoods. These actions have sparked protests from the affected indigenous groups, as they have seen significant portions of their land forcibly taken over without consent. Research shows that the loss of land impacts indigenous communities not just economically but culturally and spiritually as well. The erosion of land rights leads to the dismantling of their traditional way of life, creating a critical need for stronger legal frameworks and international support to safeguard indigenous territories from powerful external actors [38].

#### 4.5.5 Politics of Representation

The politics of representation in media coverage of indigenous protests is a critical area of concern, particularly in regions like the CHT. The under-representation of indigenous struggles, as highlighted by participants, reflects broader systemic issues. One participant from hilly lands said,

> We were protesting against the removal of graffiti that highlighted our struggles, including issues with militarization and land rights in the Chittagong Hill Tracts. Our demands were focused on indigenous rights and ending discrimination but some reports highlighted our protests as a demand for the reinstatement of the 5% quota in government jobs.

Another participant said on the same topic,

> Local media did not even mention the precursor of the protest - that the graffiti was erased.

This selective framing not only diminishes the legitimacy of the protesters' claims but also reinforces stereotypes that may further marginalize these groups. Another participant from flatland area said,

> Our protests in Netrokona focused on demands for constitutional recognition and the creation of a Ministry for lowland indigenous communities, similar to what exists for hilly regions. However, media coverage largely framed these protests as a call for security and protection, a narrative that aligns with the broader context of political shifts.

This selective portrayal is frustrating for many participants, as it undermines their fight for recognition and rights, distorting the purpose of their protests into a security issue. The distortion of protest narratives contributes to a cycle of invisibility, where the genuine needs and aspirations of indigenous communities are overlooked or misunderstood by both the media and policymakers [62].

### 4.6 Stereotypes that media perpetuates

Media representation plays a crucial role in either reinforcing or challenging these stereotypes. We asked the participants, "Does the media promote stereotypes about you?". All participants replied in the affirmative. This highlights a pervasive issue that affects their identities, livelihoods, and social relations.

#### 4.6.1 Food habit

The portrayal of indigenous food habits in media often reinforces stereotypes that misrepresents the diverse culinary practices within these communities. A participant discussed on the stereotype on Food Habit.

> I was having a regular lunch and someone said "You guys eat this as well?" It is as if our food habits are different from other people in Bangladesh.

The participant's experience of being questioned about eating habits reflects a common stereotype that limits the dietary practices of indigenous peoples to exotic dishes only. Another participant added,

> Indigenous people eat snakes, frogs and everything else.

This statement underscores another stereotype that portrays indigenous peoples as having unconventional or "primitive" eating habits, which can perpetuate harmful misconceptions about their cultures and lifestyles.

#### 4.6.2 Racism

The following statement captures the experience of indigenous individuals from the Chittagong Hill Tracts (CHT) and flatlands in Bangladesh as they encounter racial and indigenous stereotypes outside their native regions.

> If we go out of the CHT, we start facing all sorts of stereotypes. Some call us "Hey, Korean!" or "Hey, Chinese!".

The misidentification of indigenous people as "Korean" or "Chinese" highlights the superficiality of racial stereotypes that fail to recognize the unique cultural identities of these communities. Scholars have argued that racial stereotypes serve to simplify complex identities, reducing individuals to mere caricatures based on their physical appearance [39]. Continuous exposure to derogatory labels can lead to internalized racism, where individuals adopt the negative perceptions held by the broader society. Research indicates that this can contribute to a diminished sense of cultural pride and identity among indigenous youth [36].

#### 4.6.3 Extremism

The portrayal of indigenous communities in the CHT as extremists reflects a broader pattern of misrepresentation in media narratives. This framing not only perpetuates stereotypes but also marginalizes the genuine struggles faced by these communities.

> "CHT wants to be independent of Bangladesh" is a common stereotype. Any news related to this is religiously covered (e.g. Kuki-Chin [48]), But this sentiment is not common among CHT people.

And the following statement reveals significant insights into how media narratives shape public perceptions of indigenous communities and contribute to the reinforcement of stereotypes regarding extremism.

> Where there is even a small noise in Roma (where Kuki-Chin supposedly resides), all the news media covers it strongly. But the news about all the sufferings indigenous people face don't see the light.

Media attention tends to gravitate towards incidents that reinforce existing stereotypes of extremism, such as conflicts involving the Kuki-Chin community [48]. This selective coverage can create an exaggerated perception of violence and instability in the region. The media's portrayal of indigenous peoples as extremists not only misrepresents their realities but can also lead to societal stigmatization. This stigmatization can manifest in various forms, such as social exclusion and discrimination, limiting the communities' ability to advocate for their rights.

The collective feeling of invisibility among participants and their corresponding negative experiences underscore a critical need for more equitable media practices that accurately reflect the diverse realities of these marginalized communities.

### 4.7 Recommendations to news media

The participants made the following recommendations for news media. We phrase it from the participant's perspective for better delivery of their suggestions.

**Represent Education & Health Sector as well:** While culture and tourism are covered regularly, news media should expand their coverage to include critical issues such as education and healthcare. For instance, it is alarming that 60% of students in the CHT drop out of school after primary education [19]. Highlighting these educational challenges and their impacts on indigenous communities is essential for raising public awareness and prompting action. Media coverage should reflect the full spectrum of indigenous experiences, emphasizing not only cultural richness but also the pressing issues that affect their lives.

**Remove Barriers to Coverage:** To enhance the visibility and accuracy of news coverage about indigenous communities, it is essential to dismantle the barriers that restrict media reporting. This includes advocating for freedom of speech and ensuring that policies are in place to protect journalists and their ability to report on issues affecting these communities without fear of retribution. Establishing independent oversight mechanisms can help monitor and promote fair coverage, ensuring that indigenous voices are included in the media narrative.

**Educate Journalists on Historical Context:** Journalists covering news in the Chittagong Hill Tracts (CHT) should prioritize educating themselves about the region's history, cultural dynamics, and current socio-political situations. Understanding these contexts is crucial for accurate reporting. Ignorance of local issues and histories is not a valid excuse for misrepresentation. Training programs and workshops can be implemented to provide journalists with the necessary knowledge and tools to report responsibly and sensitively.

## 5 Conclusion and Future Works

Indigenous communities in Bangladesh face considerable challenges that threaten their fundamental rights, cultural identities, and livelihoods. This paper employs quantitative methods to identify articles about indigenous people and applies topic modeling and large language models (LLMs) to analyze genre distribution, sentiment, and media influence. These findings are validated through a qualitative approach, specifically through focus group discussions (FGD), which also shed light on critical issues affecting indigenous communities that the media often overlooks. The study reveals pervasive discrimination in media representation of indigenous people, with case studies highlighting severe human rights violations, such as restrictions on daily activities, surveillance, land appropriation for development projects, and consistent misrepresentation in media narratives. The FGD further reveal the profound impact of these discriminatory representations, including harmful stereotypes and mischaracterizations that shape perceptions among the broader Bengali population.

Recognizing the representation of indigenous peoples in media goes beyond accurate reporting; it has significant implications for their social, economic, and political realities. By implementing the recommendations in this paper — broadening media coverage to include topics such as education and health, dismantling barriers to accurate reporting, and equipping journalists with historical context — we can begin to foster a more equitable narrative that amplifies indigenous voices and concerns.

Future efforts should focus on developing tools to monitor media representation, ensuring that all communities have visibility in critical societal discussions. By promoting understanding and respect for indigenous rights, we can help create a more inclusive society that values the diversity and richness of all cultures. Ultimately, a commitment to fair and balanced representation is essential to empowering indigenous communities and advancing social justice in the Chittagong Hill Tracts and beyond.

Upon request, we can share the code and curated data set used in this paper.

## Appendix A: Focus Group Discussion Questions

### Discussion 1: Representation in News

- Do you feel well-represented in news?
- Can you provide examples of news stories that you believe accurately represent your community?

- Do you feel the portrayal of your community is positive, negative, or neutral? Why?
- Are there specific aspects of your culture or community that you feel are often overlooked in news coverage?

**Discussion 2: Coverage of Critical Issues**

- Do you feel that critical news is not highlighted in the media?
- What critical issues facing your community do you believe receive inadequate coverage?
- How does the lack of coverage on these issues affect your community's perception and response?
- What changes would you like to see in the media's approach to reporting critical news?

**Discussion 3: Barriers to News Access**

- Do you feel there is a barrier that filters your news?
- What barriers do you believe exist that limit access to news relevant to your community?
- Do you feel that mainstream media adequately addresses your community's concerns?
- Are there alternative sources of news that you find more reliable or relevant? If so, why?

**Discussion 4: Stereotypes in Media**

- Do you feel the media perpetuates stereotypes about you?
- What stereotypes do you feel are commonly associated with your community in the media?
- How do these stereotypes influence public perception of your community?
- Can you share any personal experiences where you felt a stereotype was perpetuated by the media?
- How can media representation change to combat these stereotypes?

**Discussion 5: Negative Experiences with Media**

- Have you encountered any personal negative experiences related to media coverage? Please share.
- How do these experiences affect your trust in the media?
- Are there specific incidents you believe should be highlighted to raise awareness?
- What support or action would you like to see from the media in response to these experiences?

**Discussion 6: Addressing Discrimination in Media**

- What specific actions do you believe news outlets should take to improve representation?
- How can the media better engage with your community to understand its needs and concerns?
- What role do you think education and training for journalists should play in addressing discrimination?